\documentclass[preprint,showpacs,preprintnumbers,amsmath,amssymb,
superscriptaddress]{revtex4}

\usepackage{graphicx}
\usepackage{dcolumn}
\usepackage{bm}

\begin{document}


\title{Precise half-life values for two neutrino double beta decay}

\author{A.S.~Barabash} \email{barabash@itep.ru}
\affiliation{Institute of Theoretical and Experimental Physics, B.\
Cheremushkinskaya 25, 117218 Moscow, Russian Federation}

\date{\today}

\begin{abstract}
All existing ``positive'' results on two neutrino double beta decay in 
different nuclei were analyzed.  Using the procedure recommended by the 
Particle Data Group, weighted average values for half-lives of 
$^{48}$Ca, $^{76}$Ge, $^{82}$Se, $^{96}$Zr, $^{100}$Mo, $^{100}$Mo - 
$^{100}$Ru ($0^+_1$), $^{116}$Cd, $^{130}$Te, $^{150}$Nd, $^{150}$Nd - $^{150}$Sm 
($0^+_1$) and $^{238}$U were obtained. Existing geochemical data were 
analyzed and recommended values for half-lives of $^{128}$Te, $^{130}$Te 
and $^{130}$Ba are proposed. Given the measured half-life values, nuclear matrix elements 
were calculated. I recommend the use of these results as
the most currently reliable values for the half-lives and nuclear matrix elements.
\end{abstract}



\maketitle

\section{Introduction}
At present, the two neutrino double beta ($2\nu\beta\beta$) decay process 
has been detected in a total of 10 different nuclei. In $^{100}$Mo and 
$^{150}$Nd, this type of decay was also detected for the transition to
the $0^+$ excited state of the daughter nucleus. For the case of the 
$^{130}$Ba nucleus, evidence for the two neutrino double electron capture 
process was observed via a geochemical experiment.  All of these results 
were obtained in a few tens of geochemical experiments and more than thirty 
direct (counting) experiments as well as and in one radiochemical experiment. In 
direct experiments, for some nuclei, there are as many as seven independent 
positive results (e.g., $^{100}$Mo).  In some experiments, the statistical 
error does not always play the primary role in overall half-life 
uncertainties. For example, the NEMO-3 experiment with $^{100}$Mo has currently detected 
more than 219,000 $2\nu\beta\beta$ events \cite{ARN05}, which results in a value for 
the statistical error of $\sim$ 0.2\% . At the same time, the systematic 
error for many experiments on $2\nu\beta\beta$ decay  
remains quite high ($\sim 10-30\%$) and very often cannot be determined 
reliably.  As a consequence, it is frequently difficult for the 
``user'' to select the ``best'' half-life value among the 
results. Using an averaging procedure, one can produce the most 
reliable and accurate half-life values for each isotope. 

Why are accurate 
half-life periods necessary? The most important motivations 
are the following:\\
{\it 1) Nuclear spectroscopy}. Now we know that some isotopes which were earlier 
considered to be stable are not, and decay via the double beta decay processes 
with a half-life period of
 $\sim 10^{18}-10^{21}$ yr are observed. The values which are presented here should be 
introduced into the isotope table.\\
{\it 2) Nuclear matrix elements (NME)}. First, it gives the possibility to improve 
the quality of NME calculations for two neutrino double beta decay, so 
one can directly compare experimental and calculated values. Second, 
it gives the possibility to improve 
the quality of NME calculations for neutrinoless double beta decay. 
The accurate half-life values for $2\nu\beta\beta$ decay
are used to adjust the most relevant parameter of the 
quasiparticle random-phase approximation (QRPA) model, the strength of the 
particle-particle interaction $g_{pp}$ \cite{ROD06,KOR07,KOR07a,SIM08}.\\
{\it 3) Research on the single state dominance (SSD) mechanism \cite{SIM01,DOM05} and a check of 
the "bosonic" component of the neutrino hypothesis \cite{DOL05,BAR07} is possible}.

In this article, an analysis of all ``positive'' experimental
results has been performed, and averaged or recommended values for  
isotopes are presented.

The first time that this work was done was in 2001, and the 
results were presented at the International Workshop on the Calculation of Double Beta Decay 
Nuclear Matrix Elements (MEDEX'01) \cite{BAR02}. Then revised half-life values were
presented at MEDEX'05 and published in Ref. \cite{BAR06}.  
In this article, new positive results obtained since 2005 have been added and analyzed.

\section{Present experimental data}

Experimental results on $2\nu\beta\beta$ decay in different nuclei are 
presented in Table 1.  For direct experiments, the number of events 
and the signal-to-background ratio are presented.

\begin{table}
\caption{Present, ``positive'' $2\nu\beta\beta$ decay results. 
Here, N is the number of useful events, $T_{1/2}$ is a half-life, and S/B is the signal-to-background 
ratio.\\
 $^{a)}$ For $E_{2e} > 1.2$ MeV; $^{b)}$ after correction (see text); 
$^{c)}$ for the SSD mechanism; $^{d)}$ in both peaks.}
\bigskip
\label{Table1}
\begin{tabular}{|c|c|c|c|c|}
\hline
\rule[-2.5mm]{0mm}{6.5mm}
Nucleus & N & $T_{1/2}$, yr & S/B & Ref., year \\
\hline
\rule[-2mm]{0mm}{6mm}
$^{48}$Ca & $\sim 100$ & $[4.3^{+2.4}_{-1.1}(stat)\pm 1.4(syst)]\cdot 10^{19}$
  & 1/5 & \cite{BAL96}, 1996 \\
 & 5 & $4.2^{+3.3}_{-1.3}\cdot 10^{19}$ & 5/0 & \cite{BRU00}, 2000 \\
& 116 & $[4.4^{+0.5}_{-0.4}(stat)\pm 0.4(syst)\cdot 10^{19}$ & 6.8 & \cite{FLA08}, 2008 \\
\rule[-4mm]{0mm}{10mm}
 & & {\bf Average value:} $\bf 4.4^{+0.6}_{-0.5} \cdot 10^{19}$ & & \\  
          
\hline
\rule[-2mm]{0mm}{6mm}
$^{76}$Ge & $\sim 4000$ & $(0.9\pm 0.1)\cdot 10^{21}$ & $\sim 1/8$                                                        
& \cite{VAS90}, 1990 \\
& 758 & $1.1^{+0.6}_{-0.3}\cdot 10^{21}$ & $\sim 1/6$ & \cite{MIL91}, 1991 \\
& $\sim 330$ & $0.92^{+0.07}_{-0.04}\cdot 10^{21}$ & $\sim 1.2$ & \cite{AVI91}, 1991 \\
& 132 & $1.2^{+0.2}_{-0.1}\cdot 10^{21}$ & $\sim 1.4$ & \cite{AVI94}, 1994 \\
& $\sim 3000$ & $(1.45\pm 0.15)\cdot 10^{21}$ & $\sim 1.5$ & \cite{MOR99}, 1999 
\\
& $\sim 80000$ & $[1.74\pm 0.01(stat)^{+0.18}_{-0.16}(syst)]\cdot 10^{21}$ & $\sim 1.5$ 
& \cite{HM03}, 2003 \\
\rule[-4mm]{0mm}{10mm}
& & {\bf Average value:} $\bf (1.5\pm 0.1) \cdot 10^{21}$ & & \\

\hline
\rule[-2mm]{0mm}{6mm}

& 89.6 & $1.08^{+0.26}_{-0.06}\cdot 10^{20}$ & $\sim 8$ & \cite{ELL92}, 1992 \\
$^{82}$Se & 149.1 & $[0.83 \pm 0.10(stat) \pm 0.07(syst)]\cdot 10^{20}$ & 2.3 & 
\cite{ARN98}, 1998 \\
& 2750 & $[0.96 \pm 0.03(stat) \pm 0.1(syst)]\cdot 10^{20}$ & 4 & \cite{ARN05}, 2005\\ 
& & $(1.3\pm 0.05)\cdot 10^{20}$ (geochem.) & & \cite{KIR86}, 1986 \\
\rule[-4mm]{0mm}{10mm}
& & {\bf Average value:} $\bf (0.92\pm 0.07)\cdot 10^{20}$ & & \\
 
\hline
\rule[-2mm]{0mm}{6mm}
$^{96}$Zr & 26.7 & $[2.1^{+0.8}_{-0.4}(stat) \pm 0.2(syst)]\cdot 10^{19}$ & $1.9^{a)}$ 
& \cite{ARN99}, 1999 \\
& 453 & $[2.35 \pm 0.14(stat) \pm 0.19(syst)]\cdot 10^{19}$ & 1 & \cite{FLA08}, 2009\\
& & $(3.9\pm 0.9)\cdot 10^{19}$ (geochem.)& & \cite{KAW93}, 1993 \\
& & $(0.94\pm 0.32)\cdot 10^{19}$ (geochem.)& & \cite{WIE01}, 2001 \\
\rule[-4mm]{0mm}{10mm}
& & {\bf Average value:} $\bf (2.3 \pm 0.2)\cdot 10^{19}$ & & \\

\hline

\end{tabular}
\end{table}

\addtocounter{table}{-1}
\begin{table}
\caption{continued.}
\bigskip
\begin{tabular}{|c|c|c|c|c|}

\hline
\rule[-2mm]{0mm}{6mm}
$^{100}$Mo & $\sim 500$ & $11.5^{+3.0}_{-2.0}\cdot 10^{18}$ & 1/7 & 
\cite{EJI91}, 1991 \\
& 67 & $11.6^{+3.4}_{-0.8}\cdot 10^{18}$ & 7 & \cite{ELL91}, 1991 \\
& 1433 & $[7.3 \pm 0.35(stat) \pm 0.8(syst)]\cdot 10^{18b)}$ & 3 & 
\cite{DAS95}, 1995 \\
& 175 & $7.6^{+2.2}_{-1.4}\cdot 10^{18}$ & 1/2 & \cite{ALS97}, 1997 \\
& 377 & $[6.75^{+0.37}_{-0.42}(stat) \pm 0.68(syst)]\cdot 10^{18}$ & 10 & 
\cite{DES97}, 1997 \\
& 800 & $[7.2 \pm 1.1(stat) \pm 1.8(syst)]\cdot 10^{18}$ & 1/9 & 
\cite{ASH01}, 2001 \\
& 219000 & $[7.11 \pm 0.02(stat) \pm 0.54(syst)]\cdot 10^{18c)}$ & 40 & 
\cite{ARN05}, 2005\\
& & $(2.1\pm 0.3)\cdot 10^{18}$ (geochem.)& & \cite{HID04}, 2004 \\ 
\rule[-4mm]{0mm}{10mm}
& & {\bf Average value:} $\bf (7.1\pm 0.4)\cdot 10^{18}$ & & \\

\hline
$^{100}$Mo - & $133^{d)}$ & $6.1^{+1.8}_{-1.1}\cdot 10^{20}$ & 1/7 & 
\cite{BAR95}, 1995 \\
$^{100}$Ru ($0^+_1$) &  $153^{d)}$ & $[9.3^{+2.8}_{-1.7}(stat) \pm 1.4(syst)]\cdot 
10^{20}$ & 1/4 & \cite{BAR99}, 1999 \\
 & 19.5 & $[5.9^{+1.7}_{-1.1}(stat) \pm 0.6(syst)]\cdot 10^{20}$ & $\sim 8$ & 
\cite{DEB01}, 2001 \\ 
& 35.5 & $[5.5^{+1.2}_{-0.8}(stat) \pm 0.3(syst)]\cdot 10^{20}$ & $\sim 8$ & 
\cite{KID09}, 2009 \\ 
& 37.5 & $[5.7^{+1.3}_{-0.9}(stat) \pm 0.8(syst)]\cdot 10^{20}$ & $\sim 3$ & 
\cite{ARN07}, 2007 \\   
\rule[-4mm]{0mm}{10mm}
& & {\bf Average value:} $\bf 5.9^{+0.8}_{-0.6}\cdot 10^{20}$ & & \\

\hline
$^{116}$Cd& $\sim 180$ & $2.6^{+0.9}_{-0.5}\cdot 10^{19}$ & $\sim 1/4$ & 
\cite{EJI95}, 1995 \\
& 9850 & $[2.9\pm 0.06(stat)^{+0.4}_{-0.3}(syst)]\cdot 10^{19}$ & $\sim 3$ & 
\cite{DAN03}, 2003 \\
& 174.6 & $[2.9 \pm 0.3(stat) \pm 0.2(syst)]\cdot 10^{19 b)}$ & 3 & 
\cite{ARN96}, 1996 \\
& 1370 & $[2.8 \pm 0.1(stat) \pm 0.3(syst)]\cdot 10^{19 c)}$ & 7.5 & \cite{FLA08}, 2008\\
\rule[-4mm]{0mm}{10mm}
& & {\bf Average value:} $\bf (2.8 \pm 0.2)\cdot 10^{19}$ & & \\

\hline
\rule[-2mm]{0mm}{6mm}
$^{128}$Te& & $\sim 2.2\cdot 10^{24}$ (geochem.) & & \cite{MAN91}, 1991 \\
& & $(7.7\pm 0.4)\cdot 10^{24}$ (geochem.)& & \cite{BER93}, 1993 \\
& & $(2.41\pm 0.39)\cdot 10^{24}$ (geochem.)& & \cite{MES08}, 2008 \\
& & $(2.3\pm 0.3)\cdot 10^{24}$ (geochem.)& & \cite{THO08}, 2008 \\
\rule[-4mm]{0mm}{10mm}
& & {\bf Recommended value:} $\bf (1.9\pm 0.4)\cdot 10^{24}$ & & \\

\hline
\end{tabular}
\end{table}

\addtocounter{table}{-1}
\begin{table}
\caption{continued 2.}
\bigskip
\begin{tabular}{|c|c|c|c|c|}

\hline
\rule[-2mm]{0mm}{6mm}
$^{130}$Te& 260 & $[6.1 \pm 1.4(stat)^{+2.9}_{-3.5}(syst)]\cdot 10^{20}$ & 1/8 & \cite{ARN03}, 2003 \\
& 236 & $[6.9 \pm 0.9(stat)^{+1.0}_{-0.7}(syst)]\cdot 10^{20}$ & 1/3 & \cite{TRE09}, 2009 \\
& & $\sim 8\cdot 10^{20}$ (geochem.) & & \cite{MAN91}, 1991 \\
& & $(27\pm 1)\cdot 10^{20}$ (geochem.)& & \cite{BER93}, 1993 \\
& & $(9.0\pm 1.4)\cdot 10^{20}$ (geochem.)& & \cite{MES08}, 2008 \\
& & $(8.0\pm 1.1)\cdot 10^{20}$ (geochem.)& & \cite{THO08}, 2008 \\
\rule[-4mm]{0mm}{10mm}
& & {\bf Recommended value:} $\bf (6.8^{+1.2}_{-1.1})\cdot 10^{20}$ & & \\

\hline
\rule[-2mm]{0mm}{6mm}
$^{150}$Nd& 23 & $[18.8^{+6.9}_{-3.9}(stat) \pm 1.9(syst)]\cdot 10^{18}$ & 
1.8 & \cite{ART95}, 1995 \\
& 414 & $[6.75^{+0.37}_{-0.42}(stat) \pm 0.68(syst)]\cdot 10^{18}$ & 6 & 
\cite{DES97}, 1997 \\
& 2018 & $[9.11^{+0.25}_{-0.22}(stat) \pm 0.63(syst)]\cdot 10^{18}$ & 2.8 & \cite{ARG09}, 2009\\
\rule[-4mm]{0mm}{10mm}
& & {\bf Average value:} $\bf(8.2\pm 0.9)\cdot 10^{18}$ & & \\

\hline
\rule[-2mm]{0mm}{6mm}
$^{150}$Nd - & $177.5^{d)}$ & $[1.33^{+0.36}_{-0.23}(stat)^{+0.27}_{-0.13}(syst)]\cdot 10^{20}$ & 
1/5 & \cite{BAR09}, 2009 \\
$^{150}$Sm ($0^+_1$) & & {\bf Average value:} $\bf 1.33^{+0.45}_{-0.26}\cdot 10^{20}$ & \\ 
 
\hline
\rule[-2mm]{0mm}{6mm}
$^{238}$U& & $\bf (2.0 \pm 0.6)\cdot 10^{21}$ (radiochem.) & & \cite{TUR91}, 1991 \\
 
\hline
\rule[-2mm]{0mm}{6mm}
$^{130}$Ba &  & $\bf (2.2 \pm 0.5)\cdot 10^{21}$ (geochem.) & 
 & \cite{MES01}, 2001 \\
ECEC(2$\nu$) & & & \\ 
\hline

\end{tabular}
\end{table}

\section{Data analysis}

To obtain an average of the ensemble of available data, a standard weighted 
least-squares procedure, as recommended by the Particle Data Group 
\cite{PDG00}, was used.  The weighted average and the corresponding error 
were calculated, as follows:
\begin{equation}
\bar x\pm \delta \bar x = \sum w_ix_i/\sum w_i \pm (\sum w_i)^{-1/2} , 
\end{equation} 
where $w_i = 1/(\delta x_i)^2$.  Here, $x_i$ and $\delta x_i$ are 
the value and error reported by the i-th experiment, and 
the summations run over the N experiments.  

The next step is to calculate $\chi^2 = \sum w_i(\bar x - x_i)^2$ and 
compare it with N - 1, which is the expectation value of $\chi^2$ if the 
measurements are from a Gaussian distribution.  If $\chi^2/(N - 1)$ is 
less than or equal to 1, and there are no known problems with the data, 
then one accepts the results to be sound.  If $\chi^2/(N - 1)$ is very large ($>> 1$), 
one chooses 
not to use the average. Alternatively, one may quote the calculated 
average while making an educated guess of the error, using a conservative 
estimate designed to take into account known problems with the data.
Finally, if $\chi^2/(N - 1)$ is larger than 1, but not greatly so, it is  
still best to use the average data, but to increase the quoted error, $\delta \bar x$ 
in Equation 1, by a factor of S defined by 
\begin{equation}
S = [\chi^2/(N - 1)]^{1/2}.
\end{equation} 
For averages, the statistical and systematic errors are treated in quadrature 
and used as a  combined error $\delta x_i$. In some cases only the results 
obtained with high enough 
signal-to-background ratio were used. 

In certain cases, the experimental results have asymmetrical errors. 
In most cases, asymmetry is small and is practically absent in the final 
result. For $^{48}$Ca, $^{100}$Mo - $^{100}$Ru ($0^+_1$) and 
$^{130}$Te the average value has the "top" error slightly larger 
than the "bottom" error, as shown in the current presentation. 
The case of $^{82}$Se is discussed in Sec. III C.

\subsection{$^{48}$Ca }    
There are three independent experiments in 
which $2\nu\beta\beta$ decay of $^{48}$Ca was observed \cite{BAL96,BRU00,FLA08}. 
The results are in good agreement. The weighted average value is:
$$
T_{1/2} = 4.4^{+0.6}_{-0.5} \cdot 10^{19} \rm{yr}.
$$ 

\subsection{$^{76}$Ge } 
Considering the results of five 
experiments, a few additional comments are 
necessary, as follows:

1) The result of the Heidelberg-Moscow group has been corrected. 
Instead of the previously published value of $T_{1/2} = [1.55\pm 
0.01(stat)^{+0.19}_{-0.15}(syst)]\cdot 10^{21}$ yr \cite{KLA01}, a new 
value $T_{1/2} = [1.74\pm 0.01(stat)^{+0.18}_{-0.16}(syst)]\cdot 10^{21}$ yr
 \cite{HM03} has been presented. It is the latter value that has been used 
in our present analysis.  At the same time, using an independent analysis, 
the Moscow part of the collaboration obtained a value similar to the result 
of Ref. \cite{HM03}, namely $T_{1/2} = 
[1.78\pm 0.01(stat)^{+0.08}_{-0.10}(syst)]\cdot 10^{21}$ yr \cite{BAK03}.

2) In Ref. \cite{AVI91}, the value $T_{1/2} = 
0.92^{+0.07}_{-0.04}\cdot 10^{21}$ yr was presented. However, after a more 
careful analysis, this result has been changed to a value of 
$T_{1/2} = 1.2^{+0.2}_{-0.1}\cdot 10^{21}$ yr \cite{AVI94}, 
which was used in the analysis.

3) The results presented in Ref. \cite{VAS90} do not agree with the more 
recent experiments \cite{HM03,MOR99}.   Furthermore, the 
error presented in \cite{VAS90} appears to be too small, especially taking 
into account that the signal-to-background ratio in this 
experiment is equal to $\sim 1/8$. It has been mentioned before
\cite{BAR90} that the half-life value in this work can be $\sim 1.5-2$ 
times higher because the thickness of the dead layer in the Ge(Li) 
detectors used can be different for crystals made from enriched Ge, rather 
than natural Ge. With no uniformity of the external background 
(and this is the case!), this 
effect can have an appreciable influence on the final result.

Finally, in calculating the average, only the results of experiments 
with signal-to-background ratios greater than 1 were used (i.e., the 
results of Refs. \cite{HM03,AVI94,MOR99}). The weighted average value is:
$$
    T_{1/2} = (1.5 \pm 0.1) \cdot 10^{21} \rm{yr}.
$$ 

\subsection{$^{82}$Se}
There are three independent counting 
experiments and many geochemical measurements $(\sim 20)$ for $^{82}$Se. The geochemical 
data are neither in good agreement with each other nor in good agreement 
with the data from the direct measurements. Typically, the accuracy of 
geochemical measurements is at the level of 10\% and
sometimes even better.  Nevertheless, the possibility of existing large 
systematic errors cannot be excluded (see discussion in Ref. \cite{MAN86}). 
It is mentioned in Ref. \cite{BAR00} that if the weak interaction constant 
$G_F$ is time-dependent, then the half-life values obtained in geochemical 
experiments will depend on the age of the samples.  Thus, to obtain a 
``present'' half-life value for $^{82}$Se, only the results of the direct 
measurements \cite{ARN05,ARN98,ELL92} were used.  The result of Ref. 
\cite{ELL87} is the preliminary result of \cite{ELL92}; hence it has not
been used in our analysis. The result of work \cite{ELL92} is presented with very 
asymmetrical errors. To be more conservative only "the top" 
error in this case is used. As a result, the weighted average value is:
$$
T_{1/2} = (0.92 \pm 0.07) \cdot 10^{20} \rm{yr}.
$$ 

\subsection{$^{96}$Zr} 
There are two ``positive'' geochemical results
\cite{KAW93,WIE01} and two results from the direct experiments of NEMO-2 \cite{ARN99} and 
NEMO-3 \cite{FLA08}.  Taking into account the comment in 
Sec. III C, I use the values from Refs. \cite{ARN99,FLA08} to obtain 
a ``present'' weighted half-life value for $^{96}$Zr of: 
$$
T_{1/2} = (2.3 \pm 0.2)\cdot 10^{19} \rm{yr}.                    
$$ 

\subsection{$^{100}$Mo} 
Formally, there are seven positive 
results from direct experiments and one recent
result from a geochemical experiment. We do not consider the result 
of Ref. \cite {VAS90a} because of a potentially high background contribution 
that was 
not excluded in this experiment. In addition, we do not consider the 
preliminary result of Elliott et al. \cite{ELL91} and instead use their 
final result \cite{DES97}, plus I do not use the geochemical result 
(again, see comment in Sec. III C).  Finally, in calculating the average, 
only the results of experiments with signal-to-background
ratios greater than 1 were used (i.e., the results of Refs. 
\cite{DAS95,DES97,ARN05}).  In addition, I have used the corrected 
half-life value from Ref. \cite {DAS95}.  Thus, the original 
result was decreased by 15\% because the calculated efficiency, in the MC,
was overestimated (see Ref. \cite {VAR97}). In addition, the half-life 
value was decreased by 10\% taking into account that for  
$^{100}$Mo we have the SSD 
mechanism (see discussion in \cite {ARN04,SHI06}). The following weighted average value 
for this half-life is then obtained:
$$
T_{1/2} = (7.1 \pm 0.4)\cdot 10^{18} \rm{yr}.                                   
$$
In the framework of the high state dominance (HSD) mechanism (see \cite{SIM01,DOM05}) the following 
average value was obtained, $T_{1/2} = (7.6 \pm 0.4)\cdot 10^{18}$ yr. 

\subsection{$^{100}$Mo - $^{100}$Ru ($0^+_1$; 1130.29 keV)} 

The 
transition to the $0^+$ excited state of $^{100}$Ru was detected in five 
independent experiments.  The results are in good agreement, and the 
weighted average for the half-life using the results from \cite{BAR95,BAR99,KID09,ARN07} is:
$$
T_{1/2} = 5.9^{+0.8}_{-0.6} \cdot 10^{20} \rm{yr} .
$$                                   
The result from \cite{DEB01} was not used here because I considered the result from \cite{KID09}
as the final result of the TUNL-ITEP experiment.

\subsection{$^{116}$Cd} 
There are four independent ``positive'' 
results \cite{FLA08,EJI95,DAN03,ARN96} that are in good agreement with each other when taking into 
account the corresponding error bars.  Again, I use here the corrected 
result for the half-life value from Ref. \cite{ARN96}.  The original 
half-life value was decreased by $\sim$ 25\% (see remark in Sec. III E). The 
weighted average value for the SSD mechanism is: 
$$          
T_{1/2} = (2.8 \pm 0.2)\cdot 10^{19} \rm{yr}.
$$ 
If the HSD mechanism is realized, then
the adjusted half-life value is $T_{1/2} = (3.0 \pm 0.2)\cdot 10^{19}$ yr. 
This is because of different single electron energy spectra for different mechanisms.
And experimental threshold in two most accurate 
experiments \cite{FLA08,ARN96} ($\sim$ 200 keV) leads to different efficiency to detect
$2\nu\beta\beta$ events.

\subsection{$^{128}$Te and $^{130}$Te} 
For a long time, there were only geochemical 
data for these isotopes. Although the half-life 
ratio for these isotopes has been obtained with good accuracy $(\sim 3\%)$ 
\cite{BER93}, the absolute values for $T_{1/2}$ of each nuclei 
are different from one experiment to the next.  One group of authors 
\cite{MAN91,TAK66,TAK96} gives $T_{1/2} \approx 0.8\cdot 10^{21}$ yr  
for $^{130}$Te and $T_{1/2} \approx  2\cdot 10^{24}$ yr for $^{128}$Te, 
whereas the next groups \cite{KIR86,BER93} claims $T_{1/2} \approx 
(2.5-2.7)\cdot 10^{21}$ yr and  $T_{1/2} \approx 7.7\cdot 10^{24}$ yr, 
respectively. Furthermore, as a rule, experiments with ``young'' 
samples ($\sim 100$ million years) give results of the half-life value of 
$^{130}$Te in the range of $\sim (0.7-0.9)\cdot 10^{21}$ yr,
while ``old'' samples ($> 1$ billion years) have half-life values in the
range of $\sim (2.5-2.7)\cdot 10^{21}$ yr. 
It has even been assumed that the difference in half-life values could be 
connected to a variation of the weak interaction constant $G_F$ with 
time \cite{BAR00}.

One can estimate the absolute half-life values for $^{130}$Te 
and $^{128}$Te using only very well-known ratios from geochemical 
measurements and the ``present'' half-life value of $^{82}$Se (see 
Sec. III C).  The first ratio \cite{BER93} is 
given by $T_{1/2}(^{130}{\rm Te})/T_{1/2}(^{128}{\rm Te}) = 
(3.52 \pm 0.11)\cdot 10^{-4}$ , while the second is 
 $T_{1/2}(^{130}{\rm Te})/T_{1/2}(^{82}{\rm Se}) = 9.9 \pm 1.5$. 
This second value is the weighted average of three experiments 
with minerals containing the elements Te and Se yield: $7.3 \pm 0.9$ 
\cite{LIN86}, $12.5 \pm 0.9$ \cite{KIR86} and $10 \pm 2$ \cite{SRI73}. 
It is significant that the gas retention age problem has no effect on 
the half-life ratio in this case. Using the ``present'' $^{82}$Se half-life 
value of $T_{1/2} = (0.92 \pm 0.07)\cdot 10^{20}$ y and the value $9.9 \pm 
1.5$ for the $T_{1/2}(^{130}{\rm Te})/T_{1/2}(^{82}{\rm Se})$ ratio, one 
obtains the half-life value for $^{130}$Te:
$$          
T_{1/2} = (9.1 \pm 2.1)\cdot 10^{20} \rm{yr}.
$$ 

Using $T_{1/2}(^{130}{\rm Te})/T_{1/2}(^{128}{\rm Te}) = 
(3.52 \pm 0.11)\cdot 10^{-4}$ 
\cite{BER93}, one obtains the half-life value for $^{128}$Te of
$$          
T_{1/2} = (2.6 \pm 0.6)\cdot 10^{24} \rm{yr}.
$$ 
Recently it was argued that "short "half-lives are more likely to be correct \cite{MES08,THO08}.
Using different "young" mineral results the half-life values were estimated at 
$(9.0 \pm 1.4)\cdot 10^{20}$ yr \cite{MES08}, $(8.0 \pm 1.1)\cdot 10^{20}$ yr \cite{THO08} 
for $^{130}$Te and $(2.41 \pm 0.39)\cdot 10^{24}$ yr \cite{MES08}, $(2.3 \pm 0.3)\cdot 10^{24}$ yr \cite{THO08} 
for $^{128}$Te, corresponding to the observed $T_{1/2}(^{130}{\rm Te})/T_{1/2}(^{128}{\rm Te})$
ratio.

The first sound indication of a positive result for $^{130}$Te in a direct experiment was obtained in 
\cite{ARN03}. A result with greater accuracy was obtained recently in the NEMO-3 experiment \cite{TRE09}.
These results are in good agreement, and the weighted average for the half-life is
$$
T_{1/2} = (6.8^{+1.2}_{-1.1})\cdot 10^{20} \rm{yr}.
$$ 
Now, using the $T_{1/2}(^{130}{\rm Te})/T_{1/2}(^{128}{\rm Te})$
ratio, one can obtain a half-life value for $^{128}$Te,
$$
T_{1/2} = (1.9 \pm 0.4)\cdot 10^{24} \rm{yr}.
$$  
We recommend the use of these last two results as the best "present" half-life values for 
$^{130}$Te and $^{128}$Te, respectively.

\subsection{$^{150}$Nd}

This half-life value was measured in three 
independent experiments \cite{ART95,DES97,ARG09}. The most accurate value was obtained in Ref. 
\cite{ARG09}. This value is higher than in Ref. \cite{DES97} and lower than in Ref. \cite{ART95} 
($\sim 3\sigma$ and $\sim 2\sigma$ differences, respectively). Using Equation 1, and three 
existing values, one 
obtains $T_{1/2} = (8.2 \pm 0.5)\cdot 10^{18}$ yr.  Taking into account 
the fact that $\chi^2 > 1$ and S = 1.89 (see Equation 2) we then obtain:
$$
T_{1/2} = (8.2 \pm 0.9)\cdot 10^{18} \rm{yr}.
$$ 

\subsection{$^{150}$Nd - $^{150}$Sm ($0^+_1$; 740.4 keV)}

There is 
only one positive result from a direct (counting) experiment \cite{BAR09}:
$$          
T_{1/2} = [1.33^{+0.36}_{-0.23}(stat)^{+0.27}_{-0.13}(syst)]\cdot 10^{20} \rm{yr}.
$$ 
The preliminary result of this work was published in \cite{BAR04}.
\subsection{$^{238}$U}  
There is again only one positive result, but this time from 
a radiochemical experiment \cite{TUR91}:
$$          
T_{1/2} = (2.0 \pm 0.6)\cdot 10^{21} \rm{yr}.
$$ 

\subsection{$^{130}$Ba (ECEC)}

Here the only positive result is 
from a geochemical experiment \cite{MES01}:
$$          
T_{1/2} = (2.2 \pm 0.5)\cdot 10^{21} \rm{yr}.
$$

In geochemical experiments it is not possible to recognise  
the different modes. But I believe this value is for the ECEC(2$\nu$) 
process because other modes are strongly suppressed (see, for example, 
estimations in \cite{DOM05,SIN07}). 

In fact, the first indication of a "positive" result for $^{130}$Ba was obtained in 
Ref. \cite{BAR96} ($T_{1/2} = 2.1^{+3.0}_{-0.8}\cdot 10^{21}$ yr) but has not been
seriously taken into account.

\section{NME values for two neutrino double beta decay}

A summary of the half-life values are presented in Table II. Using the relation 
$T_{1/2}^{-1} = G\cdot(M^{2\nu})^2$, where $G$ is the phase space factor and $M^{2\nu}$ is the nuclear 
matrix element, one can calculate $M^{2\nu}$ values for all the above mentioned isotopes. 
The results of these calculations are presented 
in Table II (3-d column). To do the calculations, I used the $G$ values from Ref. \cite{SUH98} 
for all isotopes with the exception of $^{238}$U, 
for which the $G$ value from Ref. \cite{VOG87} was used. The transition of $^{100}$Mo to the $0^+_1$ 
excited state of $^{100}$Ru used the value $G = 1.64\cdot10^{-19} yr^{-1}$ \cite{SUH09}.
Recollect that $G$ is in units of yr$^{-1}$ 
given for $g_A$ = 1.254 and
$M^{2\nu}$ is scaled by the electron rest mass. One can 
see that we now have $M^{2\nu}$ with an accuracy of
$\sim 3-14 \%$. Here it is easily noticed that the $G$ value was calculated by different authors 
(see Ref. \cite{SUH98}, Ref. \cite{DOI85}, Ref. \cite{VOG87} and Ref. \cite{SIM06}). All these results are 
in good agreement for the majority of isotopes with differences less than $1\%$. The exception 
being $^{96}$Zr with a difference 
of $\sim 6\%$; $^{100}$Mo ($\sim 6\%$); and 
$^{116}$Cd ($\sim 8\%$). One can consider these differences as systematic 
errors in the $G$ value. It means that the accuracy for $M^{2\nu}$ 
for these three isotopes is limited to the accuracy of $G$ and is at present on the level of 
$\sim 4-6\%$. It is possible in the future that 
the $G$ calculations for these three isotopes will be improved.

\section{Conclusion}

In summary, all ``positive'' $2\nu\beta\beta$-decay results were analyzed, 
and average values for half-lives were calculated. For the cases of 
$^{128}$Te and $^{130}$Te, the so-called ``recommended'' values 
have been proposed. Using these half-life values, NMEs for two neutrino double beta decay 
were obtained.
A summary is collected in Table II. 
I strongly recommend the use of these values as 
presently the most reliable. 

Notice that the accurate half-life (or $M^{2\nu}$) values for $2\nu\beta\beta$ decay
could be used to adjust the most relevant parameter of the 
quasiparticle random-phase approximation (QRPA) model, the strength of the 
particle-particle interaction $g_{pp}$. It will give the possibility to improve 
the quality of NME calculations for neutrinoless double beta decay and, finally,
to improve the quality of neutrino mass $\langle m_{\nu} \rangle$ estimations.

\begin{table}[ht]
\label{Table1}
\caption{Half-life and nuclear matrix element values for two neutrino double beta decay 
(see Sec. IV).}
\vspace{0.5cm}
\begin{center}
\begin{tabular}{ccc}
\hline
Isotope & $T_{1/2}(2\nu)$, yr & $M^{2\nu}$ \\
\hline
$^{48}$Ca & $4.4^{+0.6}_{-0.5}\cdot10^{19}$ & $0.0238^{+0.0015}_{-0.0017}$ \\
$^{76}$Ge & $(1.5 \pm 0.1)\cdot10^{21}$ & $0.0716^{+0.0025}_{-0.0023}$ \\
$^{82}$Se & $(0.92 \pm 0.07)\cdot10^{20}$ & $0.0503^{+0.0020}_{-0.0018}$ \\
$^{96}$Zr & $(2.3 \pm 0.2)\cdot10^{19}$ & $0.0491^{+0.0023}_{-0.0020}$ \\
$^{100}$Mo & $(7.1 \pm 0.4)\cdot10^{18}$ & $0.1258^{+0.0037}_{-0.0034}$ \\
$^{100}$Mo-$^{100}$Ru$(0^{+}_{1})$ & $5.9^{+0.8}_{-0.6}\cdot10^{20}$ 
& $0.1017^{+0.0056}_{-0.0063}$ \\
$^{116}$Cd & $(2.8 \pm 0.2)\cdot10^{19}$ & $0.0695^{+0.0025}_{-0.0024}$ \\
$^{128}$Te & $(1.9 \pm 0.4)\cdot10^{24}$ & $0.0249^{+0.0031}_{-0.0023}$ \\
$^{130}$Te & $(6.8^{+1.2}_{-1.1})\cdot10^{20}$ & $0.0175^{+0.0016}_{-0.0014}$ \\
$^{150}$Nd & $(8.2 \pm 0.9)\cdot10^{18}$ & $0.0320^{+0.0018}_{-0.0017}$ \\
$^{150}$Nd-$^{150}$Sm$(0^{+}_{1})$ & $1.33^{+0.45}_{-0.26}\cdot10^{20}$
& $0.0250^{+0.0029}_{-0.0034}$ \\
$^{238}$U & $(2.0 \pm 0.6)\cdot10^{21}$ & $0.0271^{+0.0053}_{-0.0033}$  \\
$^{130}$Ba; ECEC(2$\nu$) & $(2.2 \pm 0.5)\cdot10^{21}$ & $0.105^{+0.014}_{-0.010}$  \\
\hline
\end{tabular}
\end{center}
\end{table}

\section*{Acknowledgements}

I am very thankful to Prof. S. Sutton for his useful remarks. 
A portion of this work was supported by grants from RFBR 
(06-02-72553, 09-02-92676).
This work was also supported by the Russian Federal Agency for Atomic Energy.






\bibliographystyle{aipproc}   

\bibliography{sample}

\IfFileExists{\jobname.bbl}{}
 {\typeout{}
  \typeout{******************************************}
  \typeout{** Please run "bibtex \jobname" to optain}
  \typeout{** the bibliography and then re-run LaTeX}
  \typeout{** twice to fix the references!}
  \typeout{******************************************}
  \typeout{}
 }

\end{document}